\documentstyle[12pt]{article}

\newcommand{\vs}[1]{\vspace{#1 mm}}

\newcommand{\beq}{\begin{equation}}
\newcommand{\eeq}{\end{equation}}
\newcommand{\beqa}{\begin{eqnarray}}
\newcommand{\eeqa}{\end{eqnarray}}
\newcommand{\nn}{\nonumber}
\newcommand{\eq}[1]{(\ref{#1})}
\newcommand{\ket}[1]{\vert\,{#1}\,\rangle}
\newcommand{\su}{{\cal U}_q (sl(2))}
\newcommand{\Db}{\bar \Delta}
\newcommand{\epsb}{\bar \epsilon}
\newcommand{\eps}{\epsilon}
\newcommand{\ra}{\rightarrow}
\newcommand{\delb}{\bar \partial}
\newcommand{\del}{\partial}
\newcommand{\NP}[1]{Nucl.~ Phys.~ {\bf #1}}
\newcommand{\PL}[1]{Phys. ~Lett.~ {\bf #1}}

\newcommand{\PR}[1]{Phys.~ Rev.~ {\bf #1}}
\newcommand{\PRL}[1]{Phys.~ Rev.~ Lett.~ {\bf #1}}

\newcommand{\MPL}[1]{Mod.~ Phys.~ Lett.~ {\bf #1}}

\newcommand{\JP}[1]{J.~ Phys.~ {\bf #1}:\  Math.~Gen.~}
\begin{document}
\topmargin 0pt
\oddsidemargin 1mm
\begin{titlepage}
\begin{flushright}
 OS-GE-40-93  \\
 hep-th 9312174 \\
\end{flushright}
\setcounter{page}{0}
\vs{15}
\begin{center}
{\large Quantum Group and $q$-Virasoro Symmetries in Fermion Systems}
\vs{20}

Haru-Tada Sato$^\dagger$
\vs{8}

{\em Department of Physics, Osaka University\\
Machikaneyama 1-16, Toyonaka 560, Japan}
\end{center}
\vs{10}

\begin{abstract}
We discuss a generalization of the quantum group $\su$ to the
$q$-Virasoro algebra in two-dimensional electrons system under
uniform magnetic field. It is shown that the integral representations
of both algebras are reduced to those in a (1+1)-dimensional
fermion. As an application of the quantum group symmetry, we discuss a
model of quantum group current on the analogy of the Hall current.
\end{abstract}

\vspace{1cm}

\noindent-----------------------------------------------------------------------
------------------\\
$^{\dagger}$ {\footnotesize Fellow of the Japan Society for the
Promotion of Science} \\
\mbox{}\hspace{0.4cm}{\footnotesize E-mail address~: hsato@phys.wani.
osaka-u.ac.jp}
\end{titlepage}
\newpage
%
%

\section{Introduction}
\indent

Quantum groups \cite{DJ,QG} and other deformed algebras have often
been studied in some relevance to quantum corrections, anisotropies,
discretizations and deformations of original symmetries.
Recently in the context of discretizations, in particular,
$q$-deformations of the Virasoro algebra have been developed
intensively \cite{saito}$-$\cite{cha}. The $q$-Virasoro algebras is
defined as two-loop Lie algebras which are connected to the lattice
deformation of the Liouville theory or to a discrete KdV system
\cite{belov}. Furthermore, one of them is represented in terms of a single
Majorana fermion \cite{hsato,cha} and it might well describe some other
fermion physics. We consider the fermion's phenomenon which possesses at
least the quantum subalgebra $\su$ accordingly.

There are two other reasons to consider a fermion system. The
investigation of $q$-deformations of the Virasoro algebra originated in
an attempt to deform conformal field theories (CFTs) and string
theories \cite{string}. In off-critical CFT, the algebra classifying
representations is no longer the Virasoro algebra, but some other
algebra. One of the typical solvable off-critical CFT models is also
a fermion system in an external magnetic field. We are thus
interested in whether the $q$-deformed Virasoro algebra is related to
the fermion system in a constant magnetic field.

Second, in nonrelativistic three dimensional system, we encounter
the interesting phenomena, called the quantum Hall effects\cite{QHE}.
This is also a phenomenon of electrons in magnetic field and the
$\su$ symmetry appears. Recently, it has been shown that utilizing the
$\su$ symmetry, various cases of diagonalizytion problem of a
two-dimensional electron are reduced to the Bethe ansatz equations
or solved in terms of the Bethe ansatz equations\cite{BA}. We then
expect some relevance to the $q$-Virasoro algebra.

In this paper, we are concerned with the $q$-Virasoro symmetry
as a generalization of the $\su$ in a charged particle system under a
constant magnetic field $B$ perpendicular to the $x$-$y$ plane with
the Hamiltonian
\beq
H={1\over 2m_e}(p-{e\over c}A)^2\,.\label{Ea}\eeq
In abstract stage, the quantum algebra $\su$
is defined by four generators $E^{+}$, $E^{-}$, and $k^{\pm1}$ which
satisfy the following commutation relations
\beq
[E^{+},E^{-}]
={ k^2 - k^{-2} \over q-q^{-1}}\,\,, \hskip 30pt
  k E^{\pm} k^{-1}=q^{\pm1}E^{\pm}\,.     \label{Eaa}\eeq
The $m$-dimensional matrix representation (spin-$j$ representation)
with $m =2j+1$ of these generators is
\beqa
&\pi(E^{+})=diag^{+}[[2j]_q,[2j-1]_q,\dots,[1]_q]\,,\nn \\
&\pi(E^{-})=diag^{-}[[1]_q,[2]_q,\dots,[2j]_q]\,, \label{Eab} \\
&\pi(k)=diag[q^j,q^{j-1},\dots,q^{-j}]\,,  \nn     \eeqa
where the notation $diag^+$ ($diag^-$) means upper (lower) diagonal
matrix and
\beq
[x]_q={q^x-q^{-x}\over q-q^{-1}}\,.  \label{Eac}\eeq
It is known that the generators are realized by the magnetic
translations $T_{\alpha}$
\beq
T_{(\alpha_1,\alpha_2)}=
exp({i\over\hbar}{\bf \alpha}\cdot{\bf \beta})\,,\label{Ead}\eeq
where
\beq
\beta_i = p_i -{e\over c}A_i - {eB\over c}\epsilon_{ij}x^j\,,
\hskip 30pt
A_i = -{1\over2}B\epsilon_{ij}x^j -\partial_i\Lambda\,,\label{Eae}\eeq
and $\epsilon_{11}=\epsilon_{22}=0$, $\epsilon_{12}=-\epsilon_{21}=1$.
The vector $\beta$ is related to the cyclotron center and the scalar
function will be fixed for later conveniences.

In sect. 2, we review the $\su$ symmetry in the one-body
system and survey matrix representations in order to have a prospect
how the $q$-Virasoro algebra appears. Sect. 3 is a short note on the
case of a many-particle system.
In sect. 4, we discuss the integral representations
of the $\su$ and the $q$-Virasoro generators in the nonrelativistic
field theory in three dimensions. We show that these representations
correspond to those in two-dimensional chiral fermion theory through a
dimensional reduction procedure. In sect.5, we illustrate a model of
quantum group currents on the analogy of the Hall current. In order to
complement the argument of sect.5, we notice the tensor product of two
spin-1 representations with $\nu=1/3$ in sect. 6. In this case, we
obtain the 'Hall' current with the filling factor $\nu=2/3$.

\setcounter{equation}{0}
\section{Quantum group symmetry}
\indent

The generators of the quantum group algebra $\su$ are realized by the
following combination of magnetic translations
\beq
E^{+}={T_{(\Delta, \Db)}-T_{(-\Delta, \Db)}
             \over q-q^{-1}} \label{Eba}\eeq
\beq
E^{-}={T_{(-\Delta, -\Db)}-T_{(\Delta, -\Db)}
              \over q-q^{-1}} \label{Ebb} \eeq
\beq
k=T_{(\Delta,0)}. \label{Ebc} \eeq
They satisfy the defining relations of the $\su$  with the
identification
\beq
 q=exp(i\Delta\Db \lambda^{-2})\enskip \label{Ebd}\eeq
where $\lambda$ is the magnetic length defined by $\sqrt{\hbar c/eB}$.
If we put
\beq\Delta={L_x\over 2j+1}\,,\hskip 30pt \Db={L_y\over 2j+1}\label{Ebe}
\eeq
and we assume that the total flux $N_s$ satisfies the condition
\beq
   N_s={L_xL_y\over 2\pi \lambda^2}\,, \label{Ebf} \eeq
the Landau states $\psi_l$ ($-j\leq l\leq j$, where $j$ is defined by
$N_s=2j+1$) behave as a spin-$j$ representation of the $\su$.
Namely, we gain the relations
\beq
E^{\pm}\psi_l=[{1\over2}\pm l]_q\psi_{l\pm1}\,,\hskip 30pt
k\psi_l=q^l\psi_l\,,\label{Ebg} \eeq
and
\beq
q=exp({2\pi i\over 2j+1})\,.\label{Ebh}\eeq
When the periodic boundary condition is imposed, $l$ takes an integer
value. On the other hand the anti-periodic case, $l$ takes a
half-integer value. Accordingly, whether $j$ is an integer or a
half-integer is determined from the periodic or anti-periodic boundary
condition. The matrix representation $\rho$ of the equations \eq{Ebg}
becomes
\beqa
&\rho(E^{+})=diag^{+}[-[j-{1\over2}]_q,-[j-{3\over2}]_q,\dots,
                       [j-{1\over2}]_q]\,,\nn \\
&\rho(E^{-})=diag^{-}[[j-{1\over2}]_q,[j-{3\over2}]_q,\dots,
                     -[j-{1\over2}]_q]\,,\label{Ebi} \\
&\rho(k)=diag[q^{j},q^{j-1},\dots,q^{-j} ]\,. \nn \eeqa
Using the formulae
\beq
[{m-2l\over2}]_q=[l]_q\,,\hskip 30pt[m-l]_q=-[l]_q\,
\hskip 25pt {\rm for}\hskip 15pt q^m=1 \label{Ebj}\eeq
we can verify that the representation \eq{Ebi} satisfies the commutation
relations \eq{Eaa} and it turns out that \eq{Ebi} coincides with the
spin-$j$ representation \eq{Eab} making use of the relations \eq{Ebj}.
The above representation $E^{\pm}$ changes $l$ into $l\pm1$, however
other cases of raising from $l$ to $l\pm p$ ($p\not=1$) are also possible
(see appendix A).

When $j$ is an integer, namely periodic case, the representation
\eq{Ebi} amounts to
\beqa
&\rho(E^{+})=diag^{+}[-[1]_q,-[2]_q,\dots,-[j]_q,
                       [j]_q,[j-1]_q,\dots,[1]_q]\,,\nn \\
&\rho(E^{-})=diag^{-}[[1]_q,[2]_q,\dots,[j]_q,
                     -[j]_q,-[j-1]_q,\dots,-[1]_q]\,.\label{Ebk}\eeqa
The other case of $j$ being a half-integer (anti-periodic case)
contrastively includes a zero in the entries of $diag^{\pm}$
\beqa
&\rho(E^{+})=diag^{+}[-[1]_q,-[2]_q,\dots,-[j-{1\over2}]_q,0,
                      [j-{1\over2}]_q,\dots,[1]_q]\,,\nn\\
&\rho(E^{-})=diag^{-}[[1]_q,[2]_q,\dots,[j-{1\over2}]_q,0,
                      -[j-{1\over2}]_q,\dots,-[1]_q]\,.\label{Ebl}\eeqa
We note that $j=1/2$ case is excluded in the above representation
\eq{Ebl} or \eq{Ebi} because $\rho(E^{\pm})=0$ and $\rho(k)
=diag[q^{1/2},q^{-1/2}]$. Namely, these neither satisfy the defining
relations \eq{Eaa} of the quantum algebra $\su$ nor coincide with the
representation \eq{Eab}.

These matrix representations are related to the cyclic representation
of the Curtright-Zachos type of $q$-deformed Virasoro algebra\cite{CZ}.
This is a signature of the existence of the $q$-Virasoro algebra.
Actually, the $\su$ generators \eq{Eba} and \eq{Ebb} can be generalized
into the following operators
\beq
{\hat L}_n^{(k)}={T_{(k\Delta,n\Delta)}-T_{(-k\Delta,n\Delta)}
\over q^k-q^{-k}}\,.
\eeq
This differential operator satisfies a $q$-analogue of the Virasoro
algebra (details are in sect.4).

\setcounter{equation}{0}
\section{Laughlin-Jastrow function}
\indent

In this section, we comment that the statement of the previous section
can be straightforwardly extended to many-particle system. In the
following sections, we devote ourselves to the gauge
$\Lambda=0$ and introduce the complex notation \cite{CTZ}
\beq
z={1\over \lambda\sqrt{2}}(x+iy)\,. \label{Eca} \eeq
In this notation, the one-particle magnetic translation
\eq{Ead} becomes
\beq
T_{(\eps,\epsb)}=exp(\eps b- \epsb b^{\dagger})\enskip\label{Ecb}\eeq
where
\beq b={1\over2}{\bar z}+\partial\,, \hskip 30pt
   b^{\dagger}={1\over2}z-{\bar \partial}\,, \label{Ecc}\eeq
and $\eps$ and $\epsb$ can be regarded as mutually independent
dimensionless constants. The quantum group generators are obtained by
the naive replacement of $(\Delta,\Db)$ by $(\eps,\epsb)$ in the
eqs.\eq{Eba}-\eq{Ebc}. The commutation relations \eq{Eaa} are satisfied
identifying the deformation parameter with
\beq
q=exp(-\eps\epsb)\,.\label{Ecf}\eeq
And the lowest Landau level wavefunction
\beq
\phi_l(z)=exp(-{1\over2}z{\bar z})exp(ilz) \label{Ecd}
\eeq
satisfies the relations \eq{Ebg} and \eq{Ebh} if we choose
$\eps=L_xL_y/ (\lambda N_s)^2$ and $\epsb=-i$.

In a multi-particle system, introducing the total magnetic
translation operator defined by
\beq
T^{tot}_{(\eps,\epsb)}=
\prod_{i=1}^{N_e}T^{(i)}_{(\eps,\epsb)}(x_i,y_i)\,,\label{Eci}\eeq
the quantum group generators are similarly obtained by replacing
$T$ with $T^{tot}$ in the above argument. The commutation relations
\eq{Eaa} are checked using the multiplication formula for $T^{tot}$
\beq
T^{tot}_{(n_1,n_2)}T^{tot}_{(m_1,m_2)}
=exp\left({i\over2}a^2N_e\epsilon^{ij}n_im_j\right)
T^{tot}_{(n_1+n_2,m_1+m_2)}\,,\label{Ecj}\eeq
when identifying
\beq
q=exp(-N_e \eps\epsb)\,.\label{Eck}\eeq
In this case also, the deformation parameter $q$ is given by the
filling factor $\nu=N_e/N_s$ in the lowest Landau level;
$q=exp(2\pi i\nu)$, which is a natural generalization of \eq{Ebh}.

Let us consider whether $N_e$-body wavefunctions form a representation
basis of the $\su$. For that purpose, we deal with the Laughlin-Jastrow
wavefunctions \cite{L} multiplied by the center-of-mass factor with filling
factor $\nu=1/m$ ($m$ odd)
\beq
\psi_l=exp(il\sum_i^{N_e} z_i)
       exp[-{1\over2}\sum_{i=1}^{N_e}z_i{\bar z}_i]
       \prod_{i<j}^{N_e}(z_i-z_j )^m \,.        \label{Ecm}\eeq
Operating the generators of the quantum algebra $\su$ on the above
wavefunctions \eq{Ecm}, we easily recognize that the same relations as
\eq{Ebg} are satisfied by making use of the relation \eq{Eck}.
In this consideration, the $\su$ symmetry does not concern the interactive
part of the wavefunction $\prod(z_i-z_j)$. Only the global parts
(exponential parts) are relevant \cite{Hal} and thus we only
have to discuss one-particle system in generalization to the $q$-Virasoro
algebra.

\setcounter{equation}{0}
\section{$q$-Virasoro generators}
\indent

In this section, generalizing the integral representations of the
quantum group $\su$, we derive the $q$-Virasoro algebra \cite{saito}. The
integral representations are obtained with the course of Ref.\cite{CTZ}.
After that, we discuss the connection of the representations to those
of a relativistic chiral fermion in two dimensions.

In the first quantization of the Hamiltonian
\beq
H=2a^{\dagger}a+1\,,\label{EEa}\eeq
the wavefunctions are given by two commuting harmonic oscillators
\beq
\phi_{n,l}(z)={(b^{\dagger})^{n+l}\over \sqrt{(l+n)!}}
{(a^{\dagger})^{n}
\over{\sqrt n!}}\phi_0(z)\enskip\label{EEb}\eeq
where $\phi_0(z)$ is the vacuum satisfying $a\phi_0=b\phi_0=0$ and the
argument $z$ simply means the two dimensional coordinate $(x,y)$. The
quantum numbers $n$ and $l$ correspond to the energy and the angular
momentum respectively. In terms of the wavefunctions \eq{EEb} and the
fermion oscillators $b_k^{(n)}$ and ${b_k^{(n)}}^{\dagger}$
\beq
\{\,b_k^{(n)}\,,{b_l^{(m)}}^{\dagger}\,\}=\delta_{n,m}\delta_{k,l}\,,
\label{EEc}\eeq
the second quantized field operator is defined by
\beq
\Psi(z,t)=\sum_{n,k=0}^{\infty}b_k^{(n)}\phi_{n,k-n}(z)e^{-i(2n+1)t}
\label{EEd} \eeq
and satisfies the canonical anti-commutator
\beq
\{\,\Psi(z,t)\,,\Psi(w,t)\,\}=\delta^{(2)}(z-w)\,.\label{EEe}\eeq
When commutation relations of differential operators are given by
\beq
[\,{\hat C}_a\,,{\hat C}_b\,]=f^{abc}{\hat C}_c\,,\label{EEf}\eeq
it is easy to verify that the following expression satisfies the same
commutation relations as \eq{EEf}
\beq
C_a=\int d^2w\Psi^{\dagger}(w){\hat C}_a(w,{\bar w})\Psi(w)\,,\label{EEg}\eeq
and thus \eq{EEg} provides a integral representation of the ${\hat C}$
algebra. If ${\hat C}$ is the Hamiltonian operator \eq{EEa}, $C$
becomes the conserved Hamiltonian for the non-relativistic
Schr{\"o}dinger field theory
\beq
S=\int dtd^2w\left(i\Psi^{\dagger}{\dot\Psi}-{1\over2}(D\Psi)^{\dagger}
    (D\Psi)\right)\enskip\label{EEX}\eeq
where $D_i=\partial_i-iA_i$ ($c$=$e$=1). Furthermore, if ${\hat C}_a$
is composed of the harmonic oscillators \eq{Ecc}, which commute with
the covariant derivative $D$ and with the Hamiltonian, $C_a$
is a conserved charge.

Choosing ${\hat C}_a$ to be the $\su$ operators discussed before,
the integral representation of the $\su$ generators is thus
\beqa
&E^{+}=\int d^2w\Psi^{\dagger}(w)
       {T_{(\eps,\epsb)}-T_{(-\eps,\epsb)}\over q-q^{-1}}\Psi(w)\nn\\
&E^{-}=\int d^2w\Psi^{\dagger}(w)
 {T_{(-\eps,-\epsb)}-T_{(\eps,-\epsb)}\over q-q^{-1}}\Psi(w)\label{EEh}\\
&k= \int d^2w\Psi^{\dagger}(w)T_{(\eps,0)}\Psi(w)\,.\nn \eeqa
These generators are the nonlocal conserved charges because magnetic
translations are given by the harmonic oscillators $b$ and
$b^{\dagger}$ (see \eq{Ecc}). We notice that there is an infinite number
of nonlocal conserved charges which are the generalization of \eq{EEh}
\beq
L_n^{(k)}=\int d^2w\Psi^{\dagger}(w)
   {T_{(k\eps,n\epsb)}-T_{(-k\eps,n\epsb)}\over q^k-q^{-k}}\Psi(w)\,.
\label{EEi}\eeq
These satisfy the $q$-Virasoro algebra (B.7) without a central
extension
\beq
[L_n^{(i)},L_m^{(j)}]=\sum_{\eps=\pm 1}C^{n \hskip 8pt i}
                _{m \ \eps j}L^{(i+\eps j)}_{n+m},    \label{EEj}\eeq
where the structure constants are
\beq
C^{n  \ i}_{m \ j}
     ={[(nj-mi)/2]_q[(i+j)]_q \over [i]_q[j]_q}.\label{EEk}
\eeq
For reference, the two-dimensional integral representation of the
$q$-Virasoro algebra is in appendix B. The same algebras as $\su$
and \eq{EEj} exist in both theories, however eq.(B7) possesses a
central extension.

Let us discuss the connection between two- and three-dimensional
representations. The generator \eq{EEi} is the form of
\[
\int d^2w {\delta {\cal L}_3 \over \delta{\dot\Psi}}\delta_n^{(k)}
(w,{\bar w}) \Psi(w,t),
\]
where ${\cal L}_3$ is the Lagrangian density in \eq{EEX}.
It is therefore convenient to consider the following dimensional
reduction procedure
\beq
\int d^2w{\delta {\cal L}_3 \over \delta{\dot\Psi}(w,t)}\hskip20pt
\rightarrow \hskip 20pt
{1\over2\pi i}\oint dw{\delta {\cal L}_2 \over \delta
{\bar\partial}\psi_H(w)}
\eeq
where ${\cal L}_2$ stands for the two-dimensional Lagrangian density
of a dimensionless chiral fermion. This procedure gives the change
from a nonrelativistic fermion to relativistic one except the change of
dimensions of the fermion field. The dimensions will be counted later.
We then consider the following reduced form
\beq
L_n^{(k)} \ra {1\over 2\pi i}\oint dw {1\over2}\psi_H(w)
\delta_n^{(k)}(w) \psi_H(w), \label{qv13}
\eeq
in which $\delta_n^{(k)}(w)$ should be understood as an operator of
only holomorphic parts after some projection procedure. In this paper,
we follow the method of ref.\cite{GY} to extract holomorphic parts from
$\delta_n^{(k)}(w,{\bar w})$. After moving all $\bar w$ powers to the
left of the $w$ powers, we replace ${\bar w}\ra2\del$ and $\delb\ra0$.
For example,
\[
T_{(k\eps,n\epsb)}=
exp({k\eps\over2}{\bar w})exp(n\epsb\delb)exp(-{n\epsb\over2}w)
exp(k\eps\del)\hskip 10pt\ra \hskip 10pt
exp(-n\epsb w/2)exp(2k\eps\del-\eps\epsb kn/2).
\]
Furthermore, taking account of the coordinate transformation from
$w$-cylinder to $z$-plane
\beq
w=-{2\over\epsb}\ln z \label{qv14}
\eeq
and of the parametrization \eq{Ecf}, we get the reduced variation
operator as
\beq
\delta_n^{(k)}(z)= z^n{[kz\del_z+nk/2]_q\over[k]_q}. \label{5533}
\eeq
Finally substituting $\psi_H\ra z^{1/2}\psi(z)$ into \eq{qv13} in
order that $\psi$ has the conformal weight $1/2$, we thus obtain
the following expression up to the factor $2/\epsb$
\beq
L_n^{(k)} \ra {-1\over2\pi i}\oint dz{1\over2}\psi(z)
z^n{[kz\del_z+{k\over2}(n+1)]_q\over[k]_q} \psi(z). \label{2994}
\eeq
It is easy to show that the RHS of the above equation coincides with
the $q$-Virasoro generators of the two-dimensional fermion (B.2).
We therefore recognize that the 3-d fermion currents have some share of
the two-dimensional fermion's quantum group symmetries including the
$q$-Virasoro algebra.

\setcounter{equation}{0}
\section{Quantum Group current}
\indent

We have observed the $q$-Virasoro symmetry, however we have not
understood the situation on which the $q$-Virasoro symmetry becomes
effective. In this paper, we finally discuss a model on the analogy
of the simplest model of the Hall current \cite{Laugh,Ando}.

First we review this model in short. Let us consider the gauge
$\Lambda={1\over2}Bxy+\Phi x/2\pi$ and the
cylinder with the radius $r$ defined by $L_x=2\pi r$. $\Phi$ is now a
fixed constant which will be varied adiabatically. The eigenfunctions
of the Hamiltonian are degenerated for each Landau level
$n$
\beq
\ket{l,n}=exp\left\{i{l\over r}x-{1\over2\lambda^2}(y-y_l)^2\right\}
H_n({y-y_l \over\lambda})\enskip,
\eeq
\beq
y_l={\Phi\over2\pi rB}+\lambda^2{l\over r}\enskip,\hskip 30pt l\in{\bf Z}.
\eeq
All equations in the section 2 are not changed in this case.
All the eigenfunctions for fixed value of $n$, each state is
discriminated by the $l$, i.e., the coordinate $y_l$ on the cylinder.
When we gradually increase the value of $\Phi$ up to the flux quantum
$ch/e$ over the time interval $\Delta T$, the $y_l$ becomes to
$y_{l+1}$. Accordingly all eigenfunctions $\ket{l,n}$ are shifted each
in their turn (see Fig.1) and finally settle down to the equal set of
the eigenfunctions to the original one. If all the states are
completely filled with electrons, we observe the Hall current with the
filling factor $\nu=1$. Namely,
\beq
j_x={e\over 2\pi r\Delta T},\hskip 30pt E_y={1\over2\pi rc}{\Delta\Phi
\over\Delta T}
\eeq
and we hence see $\nu=1$ in the relation
\beq
\sigma_{xy}={j_x\over E_y}=\nu{e^2\over h}.
\eeq
Hereafter, we call this mechanism as an 'entire moving of
eigenstates' (EME) and consider the analogy of the above model.
According to the localization theory, there exists a
potential which localizes wavefunctions and each Landau level splits
into the Landau subband (Fig.2), in which unlocalized states are
surrounded by localized states. The Hall current is carried by the
electrons not in localized states but in unlocalized states \cite{Aoki}
(We can depict the Hall current as in Fig.3 in the words of EME. All the
unlocalized states are connected by the arrow lines forming a right
moving current but on the other hand the localized states form closed
loops).

We furthermore need the following three assumptions \cite{Ando} to obtain
a fractional filling. (i) The magnetic field should be so strong as
to the EME close within the same Landau subband. (ii) The cylinder should
be long enough to idealize the states near the both edges of the
cylinder, namely, we assume the situation of Fig.1 near the edges. This
condition is equivalent to confining the regions of non-vanishing potential
into the middle part of the cylinder. (iii) All the states up to the final
unlocalized state (point D in Fig.2) within the Fermi surface should be
completely filled with electrons and let the Fermi surface be in the
upper localized states or in the energy gap (see Fig.2).
Because the states between E and F does not contribute to the Hall current,
there can exist unfilled localized states as well as filled localized
ones. Hence an unfilled localized state will appear in Fig.3 and this
appearance makes the filling factor being a fractional value.

Now we can make use of the above model to construct a EME current
associated to the $q$-Virasoro symmetry. For simplicity, we confine
ourselves to examine the $\su$ subalgebra part and condier five states
$e_k$ which follows the 5-dimensional representation. According to
\eq{Eab}, $E^{+}e_k=[5-k]_qe_{k+1}$ and $E^{-}e_k=[k-1]_qe_{k-1}$ are
satisfied and we should notice that $E^{+}e_2=E^{-}e_4=0$ which forbids
the state $e_3$ to be arrived from the states $e_2$ and $e_4$ when
$q^3=1$. The state $e_3$ is isolated and is therefore regarded as an
unfilled 'localized' state. In contrast with the $e_1$ moving to the
place of $e_2$, the $e_2$ moves into the $e_5$ through other generator
of the quantum group ${\hat E}^{+}$, which will be mentioned in the next
section,
\beq
{\hat E}^{\pm}={1\over [3]_q[2]_q[1]_q}(E^{\pm})^3\,.\label{Edk}
\eeq
Then the $e_5$ inevitably goes back to the $e_4$ and the $e_4$
disappears into the vanishing potential region through the action of
${\hat E}^{+}$. Consequently, the EME is accomplished by the quantum
group action.

The occupation ratio of the 'unlocalized' states including one state,
which is a destination of the outgoing state $e_4$, amounts to
\beq
\nu={4\over 5+1}={2\over3}\enskip.
\eeq
It should be noticed that this filling ratio is consistent to the
relation \eq{Eck}. The pattern of the moving of the 'electrons' is
digested in Fig.4. We should not wonder if the dimension of the
representation does not coincide with the number $m$ in $q^m=1$
contrary to the results in previous sections. The situation like this
can be realized from a tensor product decomposition, i.e.,
$3\otimes3=5\oplus3\oplus1$, which will be discussed in the next
section.

\setcounter{equation}{0}
\section{Tensor product representation}
\indent

The operation rule of the $\su$ generators on a tensor
product representation is given by the comultiplication mapping
$\Delta$ of the Hopf algebra \cite{DJ}
\beq
\Delta(E^{\pm})=E^{\pm}\otimes k^{-1}+k\otimes E^{\pm}\,,
\hskip 30pt \Delta(k^{\pm})=k^{\pm}\otimes k^{\pm}\,.  \label{Eda} \eeq
It follows that the generators on a $N$-fold tensor product
representation become
\beq
E^{\pm}=\sum_{i=1}^{N}k_1 \dots k_{i-1} E_i^{\pm} k_{i+1}^{-1}
     \dots k_{N}^{-1}\,,\hskip 25pt k=\prod_{i=1}^{N}k_i\enskip
\label{Edb}\eeq
where $E_i^{\pm}$ and $k_i$ mean the generators on the $i$-th
representation space. And the formula \eq{Ebg} is also assumed in each
space. Let us consider the tensor product of two spin-1 representations
for simplicity
\beq
\ket{i,j}=\ket{i}\otimes\ket{j}\,.\label{Edc}\eeq
The representation basis is decomposed into three representations, i.e.
the singlet $e_9$, the triplet $(e_8,e_7,e_6)$ and the quintet
$(e_5,e_4,e_3,e_2,e_1)$;
\beqa
&\hskip 10pt
e_5=\ket{1,1}\,,\hskip 25pt e_4=q^{-1}\ket{0,1}+q\ket{1,0}\nn \\
&e_3=q^2\ket{1,-1}+q^{-2}\ket{-1,1}-(q+q^{-1})\ket{0,0} \label{Edd}\\
&e_2=q\ket{0,-1}+q^{-1}\ket{-1,0}\,,\hskip 15pt e_1=\ket{-1,-1}\,,\nn
\eeqa
\beqa
&e_8=q\ket{0,1}-q^{-1}\ket{1,0}\,,\hskip 30pt
e_6=q\ket{-1,0}-q^{-1}\ket{0,-1} \nn \\
&\hskip 25pt e_7=\ket{1,-1}-\ket{-1,1}+(q-q^{-1})\ket{0,0}\,,
\label{Ede}\eeqa
and
\beq
e_9=\ket{0,0}+q^{-1}\ket{1,-1}+q\ket{-1,1}\,. \label{Edf}\eeq
The representation matrices are
\beqa
&\rho_5(E^+)=diag^+[-a,a,-(q^2+1+q^{-1})a, (q^2+q^{-2})a]\nn\\
&\rho_5(E^-)=diag^-[(q^2+q^{-2})a,-(q^2+1+q^{-1})a,a,-a]\label{Edg} \\
&\rho_5(k)=diag[q^2,q,1,q^{-1},q^{-2}]\nn \eeqa
and
\beqa
&\rho_3(E^+)=diag^+[-a,a]\,,\hskip 30pt\rho_3(E^-)=diag^-[a,-a]\,\nn\\
&\hskip 65pt\rho_3(k)=diag[q^2,1,q^{-2}]\,,\label{Edh}\eeqa
where $a=[1/2]_q$. The representations of both \eq{Edg} and \eq{Edh} satisfy
the same commutation relations as those of original 9-dimensional
representation. When $q^3=1$, they satisfy the $\su$ commutation
relations \eq{Eaa} as mentioned in previous sections.

We now notice that some new features happen in the case $q$ being a
root of unity \cite{Lu}. For example when $q^3$=1, we can easily verify
that
\beq
 (E^{\pm})^3=0\enskip\label{Edi}\eeq
and $(E^{\pm})^3$ commute with $E^{\pm}$ and $k$. As a result, $e_2$
and $e_4$ can not be arrived from $e_3$
\beq
E^{\pm}e_3=0\,.\label{Edj}\eeq
In spite of \eq{Edi}, however, $e_2$ and $e_4$ are connected with $e_5$
and $e_1$ respectively by the operator \eq{Edk}
\beq
{\hat E}^{\pm}={(E^{\pm})^3\over [3]_q!}\,. \eeq
Namely
\beqa
&{\hat E}^{+}e_1=-ae_4\,,\hskip 30pt {\hat E}^{-}e_5=-ae_2\,,\nn\\
&{\hat E}^{-}e_4=-ae_1\,,\hskip 30pt {\hat E}^{+}e_2=-ae_5\,,\label{Edl}
\eeqa
in which the value of $a$ becomes
\beq
a=(-1)^k\,,\hskip 30pt q=exp(2\pi i{k-1\over3})\,,\hskip 30pt k=2,3\,.
\label{Edm}\eeq

As mentioned in \eq{Edj}, $e_2$ and $e_4$ can not be reached from $e_3$
and so the quantum group can not move the state $e_3$ whether it is
occupied by an localized state which does not contribute to the 'Hall'
current and that the pattern of the current is the same as Fig.4
discussed previous section.

\setcounter{equation}{0}
\section{Conclusion}
\indent

We have discussed the quantum group $\su$ structure in two-dimensional
electron systems and showed that this symmetry can be extended into
the $q$-deformation of the Virasoro algebra. These symmetries exist in
many-particle systems as the Laughlin system, however they concern only
the global translational symmetry and they are essentials in one-body
system. Nevertheless, the appearance of the $q$-Virasoro algebra is
interesting, because we have not understood the reason why it appears.
The dimensionally reduced generators satisfy the centrally extended
($c=1/2$) $q$-Virasoro algebra. In this relevance, it might be related
to the $c=1/2$ representation.

Owing to the external magnetic field, $su(2)$ subalgebra of the Virasoro
algebra is deformed to $\su$ and then the $q$-Virasoro algebra appears.
This feature matches with the picture whether the system is perturbed
by some interaction corresponds to whether the conformal symmetry
is deformed or not. These quantum group generators are given by the
nonlocal charges in a nonrelativistic fermion field theory as well as
in solvable off-critical CFT models \cite{charge}. These similarities
are also interesting features of the appearance of the $q$-Virasoro
algebra.

In closing the paper, we would like to expect that the relation of
our $q$-deformed algebras or of complete quantum Virasoro algebra to
off-critical CFTs would become clear in future. We wish to speculate that
our quantum group approach serves us with new aspects of fermion
systems, related other topics \cite{any} and the qunatum Virasoro
algebra.

\vspace{1cm}
\noindent
{\em Acknowledgments}

The author would like to thank N. Aizawa, H. Awata and
H. Shinke for valuable discussions and useful suggestions,
R.I. Nepomechie and J. Daboul for their interests, A.I. Solomon
and H. Suzuki for helpful comments.
\newpage
\setcounter{equation}{0}
\appendix
\section{The case of $p\not=1$}
\indent

In this appendix, we note the case of the raising and lowering
operators $E^{\pm}$ changing the quantum number $l$ by the number $p$.
When $p=1$, $(2j+1)$-fold degeneracy is inevitably built into the
$(2j+1)$-dimensional (spin-$j$) representation and $q$ is determined to
\eq{Ebh}. We however point out that other values of $q$ are possible
in the cases of $p\geq2$.

Let us consider the $(2j+1)$-fold degeneracy and two integers $k_1$ and
$k_2$ defined by
\beq
2j+1=k_1p+k_2\,,\hskip 30pt p>k_2\,.\label{EAa}
\eeq
Instead of \eq{Ebg}, we have the relations
\beq
E^{\pm}\psi_l=[{1\over2}\pm {l\over p}]_q\psi_{l\pm p}\,,
\hskip 30pt k\psi_l=q^{l/p}\psi_l\,,\label{EAb}
\eeq
and thus have $a+1$ ($k_1+1$)-dimensional representations ($0\leq a<k_2$);
\beqa
&\rho(E^{+})=
   diag^{+}([{-j+a\over p}+{1\over2}]_q,[{-j+a\over p}+{3\over2}]_q,
\dots,[{-j+a\over p}+{2k_1-1\over2}]_q)\,, \nn\\
&\rho(E^{-})=
   diag^{-}(-[{-j+a\over p}+{1\over2}]_q,-[{-j+a\over p}+{3\over2}]_q,
\dots,-[{-j+a\over p}+{2k_1-1\over2}]_q)\,, \label{EAc} \\
&\rho(k)=diag(q^{-(j+a)/p+k_1},q^{-(j+a)/p+k_1-1},\dots,
q^{-(j+a)/p})\,. \\ \eeqa
If we put
\beq
q=exp(2\pi i{p\over p+2j})\,, \label{EAd}
\eeq
the representation matrices become simple forms. For example, $E^{+}$
of $a=0$ representation reads
\beq
\rho(E^{+})=diag^{+}(-[1]_q,-[2]_q,\dots,-[k_1]_q)\,. \label{EAe}
\eeq
The simplest case is $k_2=1$, namely the only possible value of $a$ is
zero. Eq.\eq{EAd} becomes
\beq
q=exp({2\pi i\over k_1+1})\enskip\label{EAf}
\eeq
and the representation matrices \eq{EAc} coincides with spin-($k_1/2$)
representation with the value of \eq{EAf}. We then conclude that
when the generators label the quantum number $l$ by the $p$ ($\not=1$),
$q$ is related not to the filling factor but to the dimensions of
representations only.
\setcounter{equation}{0}
\section{Neveu-Schwarz fermion}
\indent

We show the realization of the $\su$ generators by the Neveu-Schwarz
fermion in two dimensions\cite{belov,hsato,cha};
\beq
  \psi(z)=\sum_r {b_r \over z^{r+1/2} },
    \hskip 30pt r\in{\bf Z}+{1\over 2}.            \label{EBa}\eeq
Let us define the following Fourier mode of a nonlocal current
\beq
L^{(k)}_n={-1\over2\pi i}\oint dzz^n:\psi(zq^{-k/2})
     {q^{{k\over2}z\partial}-q^{-{k\over2}z\partial}\over q^k-q^{-k}}
     \psi(z):\,\,.     \label{EBb}\eeq
The explicit form of $L^{(k)}_n$ in terms of the fermion oscillators is
\beq
L^{(k)}_n={1\over 2[k]}\sum_r[{n-2r \over 2}k]:b_r b_{n-r}: \label{EBc}
\eeq
where :${\phantom C}$: denotes the normal ordering defined by
\beq
:b_rb_s:=b_rb_s-\theta(r)\{b_r,b_s\}\,,\hskip 35pt
\{b_r,b_s\}= \delta_{r+s,0}.         \label{EBd}\eeq
The generators which satisfy the commutation relations \eq{Eaa} are
\beq
E^{+}=L^{(1)}_1\,,\hskip 30pt E^{-}=-L^{(1)}_{-1}\,,\hskip 30pt
k=q^{-L_0}\,,\label{EBe}\eeq
where $L_0$ is the zero mode of the Virasoro operator
\beq
L_n={1\over 2}\sum_r({1\over2}n-r):b_r b_{n-r}:\,. \label{EBf}\eeq
Furthermore, \eq{EBc} satisfies the $q$-Virasoro algebra with a
central extension
\beq
[L_n^{(i)},L_m^{(j)}]=\sum_{\eps=\pm 1}C^{n \hskip 8pt i}
    _{m \ \eps j}L^{(i+\eps j)}_{n+m}
    +{1\over2}C_{ij}(n)\delta_{n+m},\label{EBg}
\eeq
where $C^{n  \ i}_{m \ j}$ is given by \eq{EEk} and
\beq
C_{ij}(n)
={1\over [i][j]}\sum_{k=1}^n[{(n+1-2k)i\over 2}][{(n+1-2k)j\over 2}]\,.
\eeq


%
\pagebreak[4]
\topmargin 0pt
\oddsidemargin 5mm
\centerline{\bf FIGURE CAPTIONS}

\begin{description}
\item[{\bf Fig.1}:] The Hall current with $\nu=1$ induced by $\Delta\Phi$.
\item[{\bf Fig.2}:] Each Landau level becomes the Landau subband like
a mountain. The regions {\bf A} and {\bf C} correspond to the localized
states and the region {\bf B} to the unlocalized ones. The point {\bf F}
means the Fermi energy.
\item[{\bf Fig.3}:] In the presence of a localization potential, EME
is not like Fig.1 but like Fig.3. The closed loop corresponds to a
localized state.
\item[{\bf Fig.4}:] The pattern of the quantum group 'Hall' current
with the filling factor $\nu=2/3$. $e_3$ is a 'localized' state which
does not contribute the current.

\end{description}

\end{document}